\documentclass[final,5p,times]{elsarticle}

\usepackage[T1]{fontenc}
\usepackage[utf8]{inputenc}
\usepackage[left]{lineno}
\usepackage{amsmath,amssymb}
\usepackage{graphicx}
\usepackage{booktabs}
\usepackage{array}
\usepackage{tabularx}
\usepackage{siunitx}
\usepackage{xcolor}
\usepackage{caption}
\usepackage{microtype}
\usepackage[percent]{overpic}

\usepackage[hidelinks]{hyperref}

\makeatother
\begin{document}
\begin{frontmatter}

\title{Improved aluminum film transport properties
and enhanced TES response in athermal phonon detectors:
implications for quasiparticle dynamics}

\author[inst]{M. Mirzakhani}
\author[inst]{K. Hunter}
\author[inst]{R. Mahapatra}
\author[inst]{M. Platt}

\affiliation[inst]{
  organization={Department of Physics and Astronomy,
                Texas A\&M University},
  addressline={578 University Dr},
  city={College Station},
  postcode={77840},
  state={TX},
  country={US}
}

\begin{abstract}
Our detectors for Dark Matter and Coherent Elastic Neutrino--Nucleus Scattering (CE$\nu$NS) use photolithographically patterned superconducting athermal phonon sensors that are coupled to kg-scale absorbers of different materials (germanium, silicon, sapphire). Following a particle interaction, athermal phonons from the bulk absorber break Cooper pairs in a superconducting aluminum film that surrounds a tungsten Transition Edge Sensor (TES) in equilibrium through electrothermal feedback. This sensor design is called Quasiparticle-trap-assisted Electro-thermal-feedback Transition-edge-sensor (QET). The QET aluminum film that absorbs the athermal phonons is patterned into fins with lengths limited by the quasiparticle diffusion length in aluminum. Improving the quasiparticle diffusion length in the superconducting aluminum is an area of vigorous research to improve phonon collection efficiency and increase the signal strength for lower-threshold searches. We report significantly improved aluminum films that show systematically larger grain sizes and residual-resistivity ratio (RRR) compared to detectors we have fabricated in the past. The improved RRR and grain sizes were expected to improve the collected signal for a given energy. We report higher signal strength achieved in a detector in comparison to another detector in which all other fabrication parameters were kept identical except for the aluminum film deposition. 

\end{abstract}


\end{frontmatter}

% Place \end{document} at the end of the complete paper.
\section{Introduction}
\label{sec:introduction}

Cryogenic phonon detectors measure energy deposited in solid
absorbers by collecting phonons generated by particle
interactions. In quasiparticle-trap-assisted
electrothermal-feedback transition-edge sensors (QETs),
superconducting aluminum collectors are coupled to tungsten
transition-edge sensors (TESs) \cite{Irwin1995, CDMS-II:2009ktb, article, TESSERACT:2025odn, TESSERACT:2025tfw}. Phonons break
Cooper pairs in the aluminum, generating quasiparticles that
diffuse toward the tungsten and transfer energy to the TES \cite{PhysRevLett.20.1502, LOIDL2001440}.
This architecture increases the phonon-collection area
without increasing the tungsten sensing volume \cite{BihariPrasad:2013mfw}.

Under voltage bias, a TES operates within its superconducting
transition with negative electrothermal feedback: a temperature
increase raises its resistance and reduces Joule heating.
The resulting current changes are measured using a
superconducting quantum interference device (SQUID)
\cite{Irwin1995}. The measured response therefore depends on
both the energy delivered to the substrate and the electrical and
thermal properties of the sensor and readout system.

The overall detector performance is a combination of efficiencies of the quasiparticle elastic scattering and lifetime in aluminum, as well as the quasiparticle transmission from aluminum to tungsten. Quasiparticle transport and losses in the aluminum, together with energy transfer at the Al/W interface, influence the energy reaching the TES \cite{Yen2014}. Because our aluminum films are \SI{600}{\nano\metre}
thick, the primary limiting factors are bulk defect
scattering, grain boundaries, and interface
transparency at the TES. We utilize two characterization benchmarks to predict overall detector performance improvements, followed by full detector signal comparison at mK temperatures. The two characterization benchmarks provide information on the elastic mean free path in our thick aluminum film through measurements of grain sizes using Atomic Force Microscopy (AFM) imaging and Residual Resistivity Ratio (RRR) above the aluminum superconducting temperature. Because deposition conditions can affect film disorder, surface morphology,
and interface quality, optimizing aluminum deposition offers
a potential route to improving detector response.
%We characterize film transport using the RRR,
%\begin{equation}
%\mathrm{RRR} = \frac{\rho(T_{\mathrm{room}})} {\rho(T_0)},
%\label{eq:rrr}
%\end{equation}
%where \(T_{\mathrm{room}}\) denotes room temperature and \(T_0 \approx \SI{10}{\kelvin}\) is the low-temperature reference above the aluminum superconducting transition.
%This ratio provides a comparative measure of film transport, rather than a direct measurement of quasiparticle lifetime or collection efficiency.

To promote large columnar grain growth, we deposit
the improved aluminum film at an elevated substrate
temperature. We find a strong correlation between
RRR and grain size in aluminum films deposited
as part of a designed experimental study.
We selected the baseline aluminum film as the control
and a film with approximately twice the RRR and
substantially larger grains for the test detector.

In this work, we compare two silicon phonon detectors with
the same nominal channel layout and tungsten deposition
process, but with aluminum collectors deposited using either
an AJA or a SEGI sputtering system. We combine film-transport
and surface-morphology measurements with cryogenic electrical
characterization and pulse-spectrum analysis to examine
the association between aluminum film properties and detector
response, while performing a relative comparison between the two detectors. %while distinguishing the measured response from absolute collection efficiency.

\section{Detector Design and Fabrication}
\label{sec:fabrication}

\subsection{Silicon substrates and channel layout}

Two silicon substrates, each \SI{1}{\milli\metre} thick, were
used as the absorbers in this study. Energy deposited by
particle interactions in the silicon generates phonons,
a fraction of which is collected by the aluminum fins
coupled to the tungsten transition-edge sensors. Each detector contained three inner readout channels,
arranged symmetrically around the substrate center, and one
outer ring channel.

Before thin-film deposition, the substrates underwent
a chemical cleaning procedure. They were first rinsed
sequentially with methanol and isopropyl alcohol, followed
by spin rinsing and drying (SRD). A subsequent piranha
cleaning step was performed to further prepare the
substrate surfaces for deposition \cite{Jastram:2014pga}.

\begin{figure}[htbp]
\centering
\includegraphics[width=\linewidth]{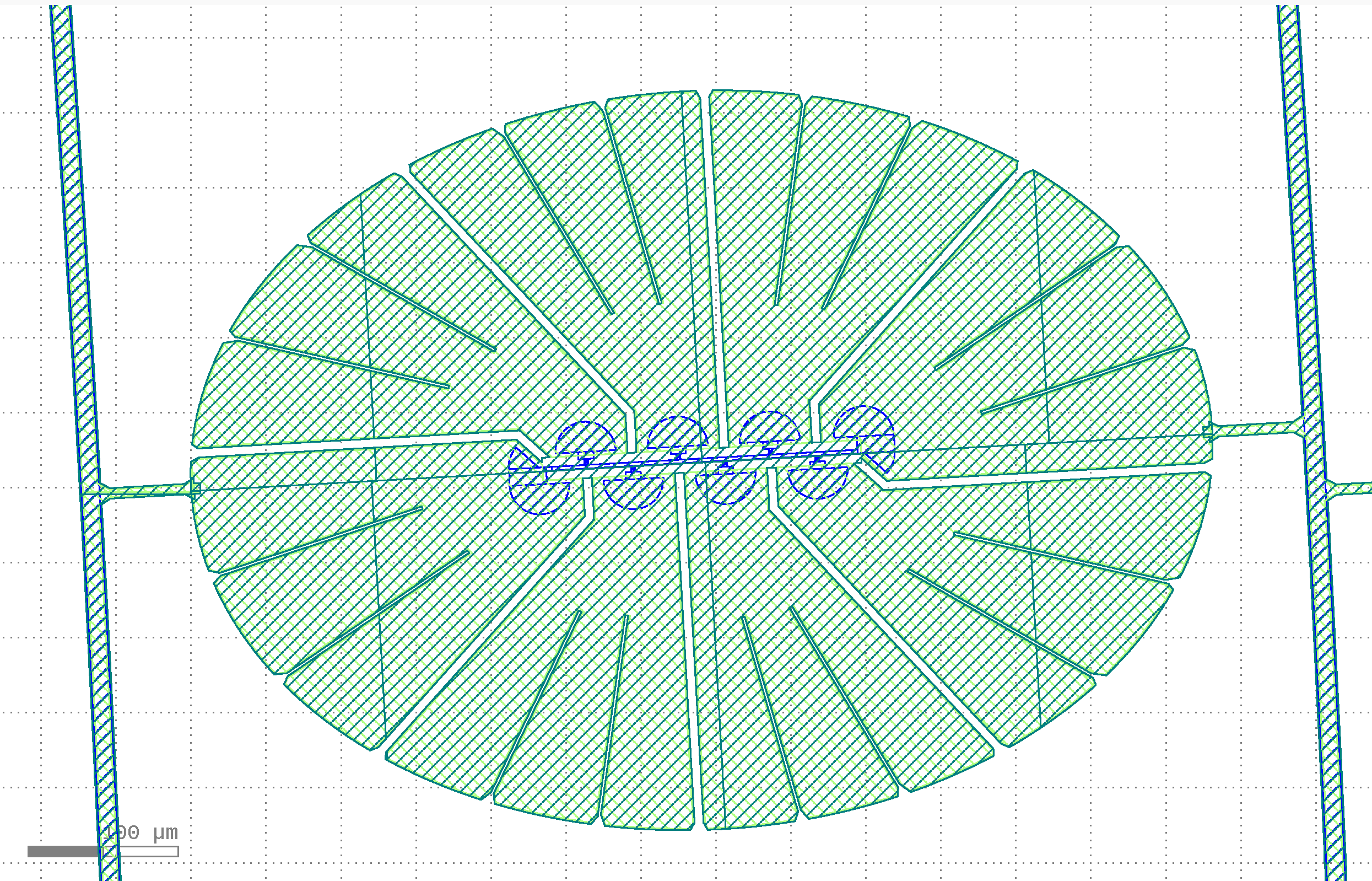}
\caption{
QET sensor layout extracted from the GDS design file.
The green regions represent the aluminum phonon-collection
fins, while the blue regions show the tungsten TES
structure, including the semicircular Al/W overlap
regions.
}
\label{fig:QET}
\end{figure}

\begin{figure}[ht!]
    \centering

    \begin{overpic}[width=\linewidth]{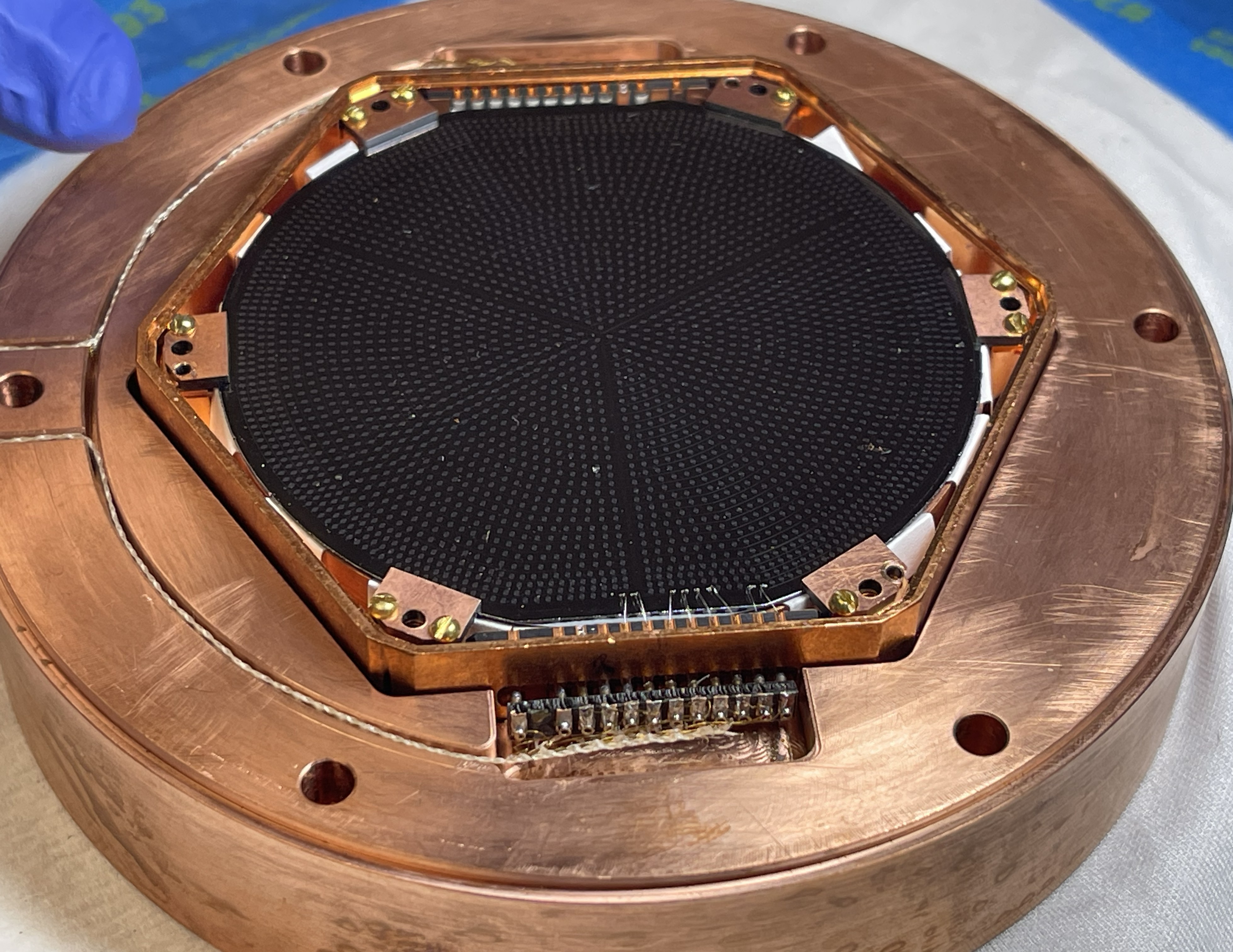}
        \put(70,0){%
            \includegraphics[width=0.30\linewidth]{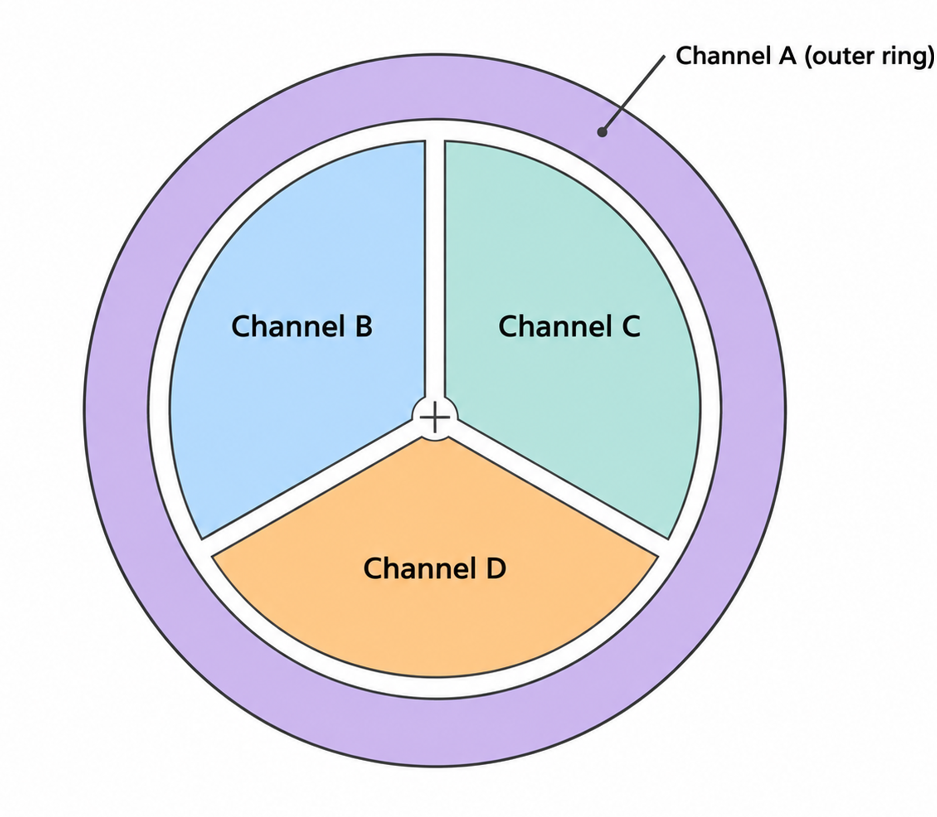}%
        }
    \end{overpic}

    \caption{
        Top-view photograph of the detector (main image)
        and schematic of the common detector layout (inset),
        showing three symmetrically arranged inner channels
        and an outer ring channel. Both detectors have the
        same nominal layout and follow the detector layout designed for the MINER experiment~\cite{MINER:2025wrk, MINER:2026sat}.}
    \label{fig:layout}
\end{figure}

The substrate dimensions, resistivity,
surface-preparation details, and aluminum collector
and tungsten TES geometries are summarized in
Tab.~\ref{tab:geometry}.
The QET sensor layout extracted from the GDS file
is shown in Fig.~\ref{fig:QET}, while a schematic
and a photograph of the detector are presented
in Fig.~\ref{fig:layout}.

\begin{table}[htbp]
\centering
\caption{Nominal detector geometry and substrate properties.}
\label{tab:geometry}
\small
\begin{tabular}{@{}p{0.52\linewidth} p{0.40\linewidth}@{}}
\toprule
Parameter & Value \\
\midrule
Absorber material & Silicon \\
Substrate diameter & \SI{76}{\milli\metre} \\
Substrate thickness & \SI{1}{\milli\metre} \\
Silicon resistivity &
\SI[inter-unit-product=\ensuremath{{}\cdot{}}]
   {>8}{\kilo\ohm\centi\metre} \\
Inner channels & B, C, and D; symmetrically arranged \\
Outer channel & A; annular geometry \\
%Instrumented channels & B and C \\
Al collector diameter & \SI{400}{\micro\metre} \\
W TES dimensions &
\(\SI{200}{\micro\metre}
  \times \SI{2}{\micro\metre}\) \\
TES elements per channel & 840 \\
Al/W overlap geometry &
\SI{40}{\micro\metre} diameter semicircle \\
Al surface coverage fraction & 40\% \\
\bottomrule
\end{tabular}
\end{table}

\subsection{Thin-film deposition and patterning}
\label{sec:deposition}

Nominally \SI{40}{\nano\metre}-thick tungsten films were
deposited on both silicon substrates using the SEGI
sputtering system at a DC power of \SI{1.4}{\kilo\watt}
and a working pressure of \(21\,\mathrm{mTorr}\).
A residual gas analyzer (RGA) monitored the chamber
gas composition. The films were then patterned by
mask-based photolithography and etched to define the
transition-edge sensor structures, followed by
photoresist stripping.

Before aluminum deposition, the patterned tungsten
surfaces of both samples underwent argon etching to
remove surface oxide and prepare the Al/W contact
regions~\cite{546406}.

Nominally \SI{600}{\nano\metre}-thick aluminum films
were subsequently deposited using either the AJA
or SEGI system. The AJA deposition used a DC power
of \SI{350}{\watt} for \SI{40}{\minute}, whereas
the SEGI deposition used \SI{1}{\kilo\watt} for
\SI{45}{\minute}. Both processes operated at a
working pressure of \(10\,\mathrm{mTorr}\).
The nominal tungsten and aluminum thicknesses
were estimated from their respective deposition
rates and durations.

A second photolithography step, aligned to the
tungsten structures, followed by aluminum etching
defined the phonon-collection fins and their overlap
regions with the TES elements. The deposition parameters are
summarized in Tab.~\ref{tab:recipes}.

\begin{table}[ht!]
\centering
\caption{
    Deposition parameters for the AJA and SEGI detectors.
    Detector labels identify the aluminum deposition system;
    tungsten was deposited using SEGI for both detectors.
}
\label{tab:recipes}
\small
\setlength{\tabcolsep}{4pt}
\renewcommand{\arraystretch}{1.1}

\begin{tabularx}{\linewidth}{
    @{}
    >{\raggedright\arraybackslash}p{0.38\linewidth}
    >{\raggedright\arraybackslash}X
    >{\raggedright\arraybackslash}X
    @{}
}
\toprule
Parameter & AJA detector & SEGI detector \\
\midrule
W deposition system & SEGI & SEGI \\
Al deposition system & AJA & SEGI \\
Sputtering mode & DC & DC \\
Tungsten purity (\%) & 99.9999 & 99.9999 \\
Aluminum purity (\%) & 99.999 & 99.999 \\
Base pressure
    & \(\sim 7 \times 10^{-8}\,\mathrm{Torr}\)
    & \(\sim 3 \times 10^{-7}\,\mathrm{Torr}\) \\
Working pressure
    & \(10\,\mathrm{mTorr}\)
    & \(10\,\mathrm{mTorr}\) \\
Al sputtering power
    & \SI{350}{\watt}
    & \SI{1}{\kilo\watt} \\
Substrate temperature
    & \SI{150}{\degreeCelsius}
    & \SI{21}{\degreeCelsius} \\
Al deposition duration
    & \SI{40}{\minute}
    & \SI{45}{\minute} \\
Nominal Al thickness
    & \SI{600}{\nano\metre}
    & \SI{600}{\nano\metre} \\
Reported Al RRR
    & \(\approx 18\)
    & \(\approx 9\) \\
%W deposition system & SEGI & SEGI \\
\bottomrule
\end{tabularx}
\end{table}

\begin{figure*}[t]
    \centering
    \includegraphics[width=\textwidth]{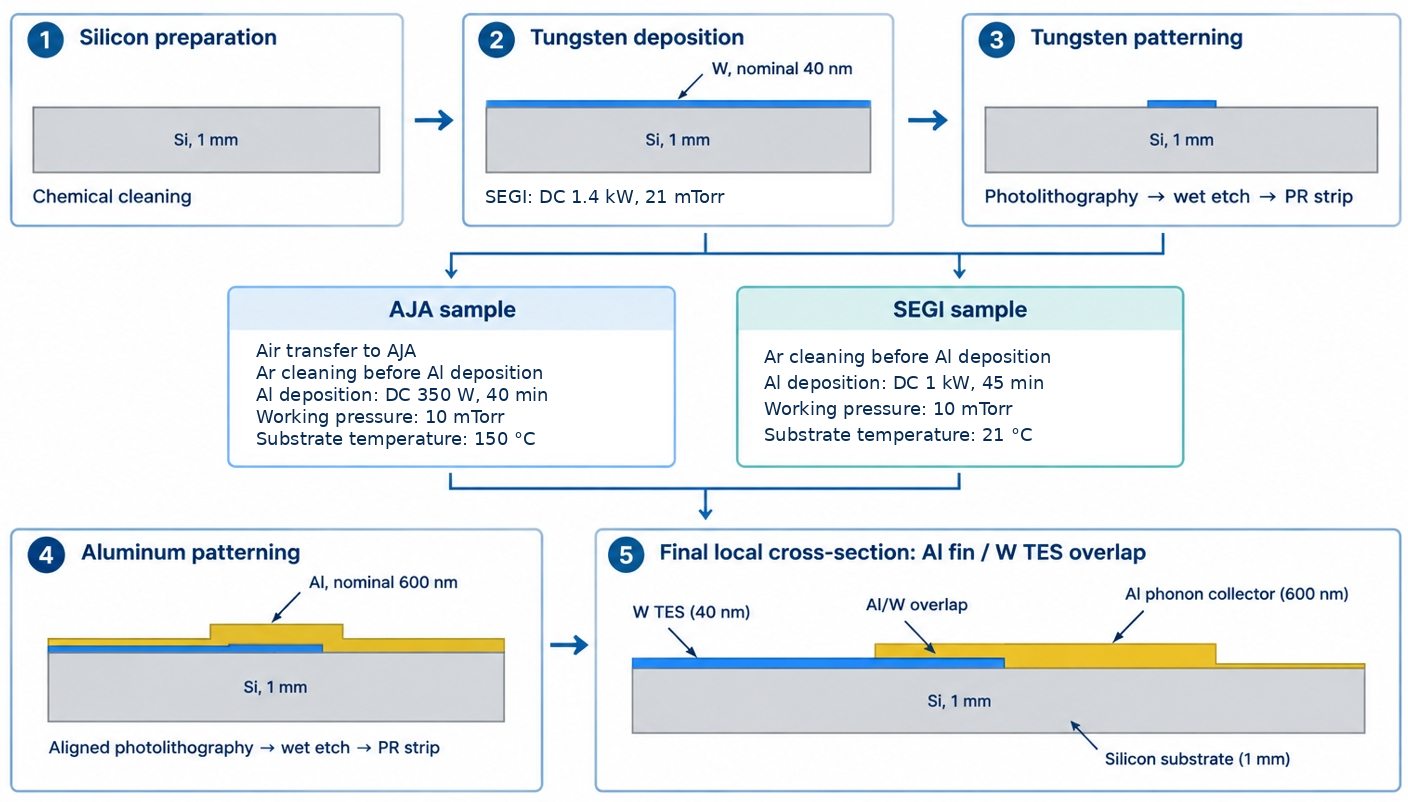}
\caption{
    Fabrication sequence and schematic cross-section
    of the Al/W QET structures. The final cross-section
    illustrates the local Al/W overlap geometry on the
    silicon substrate. The schematics are not to scale
    and were generated using an AI image-generation
    tool based on the fabrication procedure described
    in this work.
}
    \label{fig:fabrication_sequence}
\end{figure*}

\subsection{Photolithographic patterning}
\label{sec:photolithography}

Both detectors were patterned using the same nominal
design and sequential mask-based photolithography.
For each metal layer, the process comprised photoresist
coating and baking, mask exposure, development,
wet etching, and photoresist stripping.

The first step defined the tungsten TES structures.
Following argon surface cleaning and aluminum deposition
as described in Section~\ref{sec:deposition}, the second
mask was aligned to the tungsten structures to define
the aluminum phonon-collection fins and Al/W overlap
regions.

Photolithography, and etching parameters are
summarized in Tab.~\ref{tab:PR_recipes}.

\begin{table}[ht!]
\centering
\caption{
Photolithography and wet-processing parameters
used for both detectors.
}
\label{tab:PR_recipes}
\small
\begin{tabular}{@{}p{0.58\linewidth} p{0.34\linewidth}@{}}
\toprule
Parameter & Value \\
\midrule
Photoresist & S1811 \\
Spin-coating speed & 2300 rpm \\
Photoresist bake duration & \SI{5}{\minute} \\
Photoresist bake temperature & \SI{120}{\degreeCelsius} \\
Developer & MF319 \\
Development duration & \SI{90}{\second} \\
Tungsten etchant & 30\% \(\mathrm{H_2O_2}\) \\
Tungsten etching duration & \SI{7}{\minute} \\
Aluminum etchant & Al etchant \\
Aluminum etching duration & \SI{12.5}{\minute} \\
Photoresist stripper & PRS3000 \\
Photoresist stripping duration & \SI{15}{\minute} \\
\bottomrule
\end{tabular}
\end{table}

\section{Film Characterization}
\label{sec:films}

\subsection{Residual-resistivity ratio}
\label{sec:rrr}

Aluminum witness samples with an area of approximately
\SI{1}{\centi\metre\squared}, corresponding to the AJA
and SEGI deposition processes, were characterized by
four-wire resistance measurements at room temperature,
\(T_{\mathrm{room}}\), and at \(T_0 \approx
\SI{10}{\kelvin}\) in a small cryostat.

Following the definition in Eq.~\ref{eq:rrr}, the RRR
was evaluated from the measured resistances as
\begin{equation}
\mathrm{RRR}
=
\frac{R(T_{\mathrm{room}})}{R(T_0)}.
\label{eq:rrr}
\end{equation}
The four-wire configuration minimized contributions
from leads and contacts. Because the measurement
geometry was unchanged between temperatures, the
resistance ratio represented the corresponding
resistivity ratio.

The low-temperature reference was chosen to approximate
the residual-resistance regime. Thus, the reported
values represent room-temperature-to-\SI{10}{\kelvin}
resistance ratios rather than extrapolated
zero-temperature values.

The AJA and SEGI witness samples exhibited RRR values
of approximately 18 and 9, respectively, providing
a comparative measure of aluminum film transport
properties for the two deposition processes.

\subsection{Atomic force microscopy}
\label{sec:afm}

Atomic Force Microscopy (AFM) was used to compare
the grain size of separate, nominally
\SI{300}{\nano\metre}-thick aluminum characterization
films deposited using the AJA and SEGI systems.
Each sample was prepared using the corresponding
detector deposition recipe, with only the deposition
duration adjusted to obtain the lower nominal thickness.
Because the detector collectors were nominally
\SI{600}{\nano\metre} thick, these measurements
provide supplementary characterization of the
deposition processes rather than direct measurements
of detector-film morphology.

Both samples were measured using a DIM 4000 scanner
with an SG head, as identified in the acquisition
metadata. Each height image covered a
\(20\times20\,\mu\mathrm{m}^{2}\) region with
256 samples per scan line. Height maps and
three-dimensional renderings were analyzed using
NanoScope Analysis.

Surface roughness was quantified by the
root-mean-square height variation,
\begin{equation}
R_q =
\sqrt{\frac{1}{N}\sum_{i=1}^{N}(z_i-\bar{z})^2},
\label{eq:roughness}
\end{equation}
and the arithmetic mean height deviation,
\begin{equation}
R_a =
\frac{1}{N}\sum_{i=1}^{N}|z_i-\bar{z}|,
\label{eq:arithmetic_roughness}
\end{equation}
where \(z_i\) is the height at pixel \(i\),
\(\bar{z}\) is the mean height, and \(N\) is the
number of pixels included in the analysis.

\begin{figure}[ht!]
\centering
    \includegraphics[width=\linewidth]{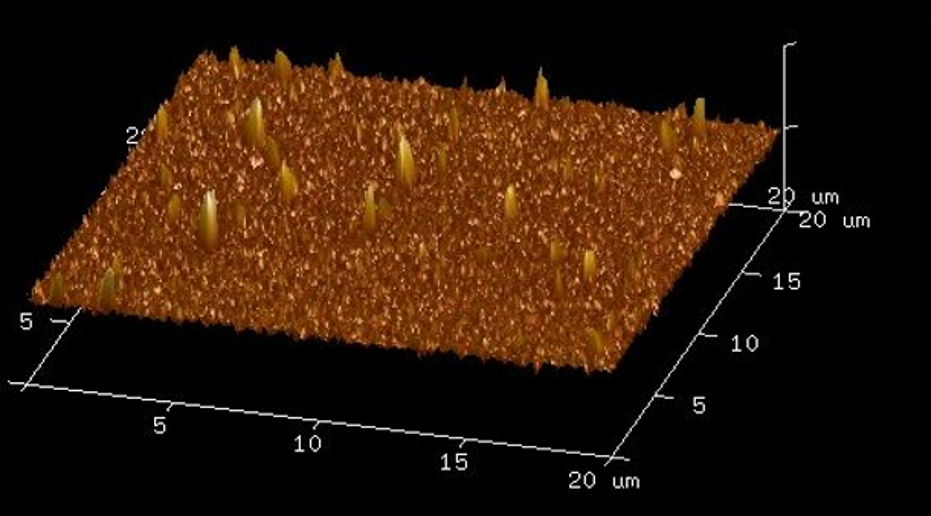} \\
    \includegraphics[width=\linewidth]{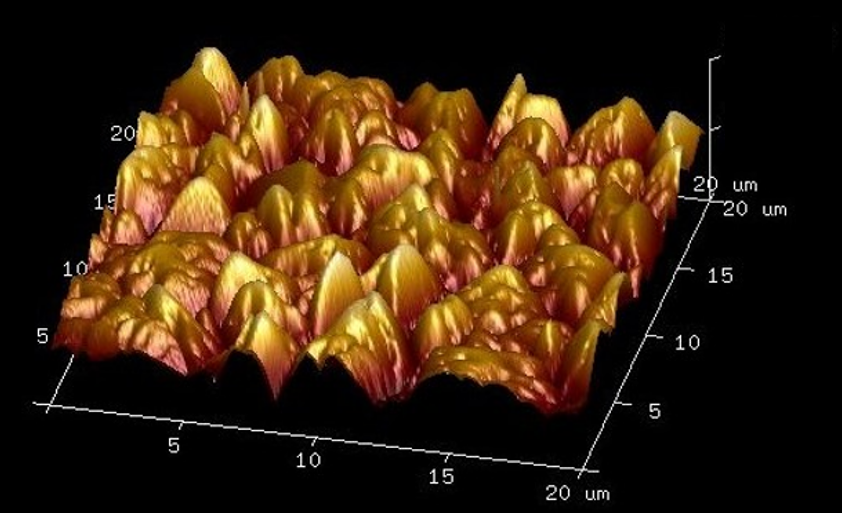}\\
\caption{
Three-dimensional AFM renderings of nominally
\SI{300}{\nano\metre}-thick aluminum characterization
films deposited using SEGI (top) and AJA (bottom).
The vertical axes describe surface-height variations,
not film thickness. These characterization films
are distinct from the nominally
\SI{600}{\nano\metre}-thick detector aluminum films.
}
\label{fig:afm_characterization}
\end{figure}

For the analyzed regions, the AJA film exhibited
\(R_q=\SI{132}{\nano\metre}\) and
\(R_a=\SI{112}{\nano\metre}\), compared with
\(R_q=\SI{9.54}{\nano\metre}\) and
\(R_a=\SI{5.60}{\nano\metre}\) for the SEGI film.
Both roughness metrics were therefore substantially
larger for the AJA sample. The results and
three-dimensional renderings are presented in
Table~\ref{tab:afm_results} and
Fig.~\ref{fig:afm_characterization}, respectively.

\begin{table}[htbp]
\centering
\caption{
AFM roughness of the nominally
\SI{300}{\nano\metre}-thick aluminum characterization
films. Each scan covered a
\(20\times20\,\mu\mathrm{m}^{2}\) region.
}
\label{tab:afm_results}
\small
\begin{tabular}{@{}lll@{}}
\toprule
Parameter & AJA & SEGI \\
\midrule
Samples per scan line & 256 & 256 \\
Projected scan area
& \SI{400}{\micro\metre\squared}
& \SI{400}{\micro\metre\squared} \\
RMS roughness, $R_q$
& \SI{132}{\nano\metre}
& \SI{9.54}{\nano\metre} \\
Arithmetic roughness, $R_a$
& \SI{112}{\nano\metre}
& \SI{5.60}{\nano\metre} \\
\bottomrule
\end{tabular}
\end{table}

\section{Cryogenic Assembly and Readout}
\label{sec:setup}

\subsection{Copper tower and shielding}
\label{sec:copper_tower}

The detectors were mounted in a copper tower with
an approximate wall thickness of \SI{2}{\milli\metre},
with the AJA detector below the SEGI detector.
A copper plate approximately \SI{1}{\centi\metre}
thick was placed between them. Together with the
copper enclosure, this plate was intended to reduce
infrared exposure and direct optical coupling to
the detectors. Gaps and seams between the stacked
detector holders were covered with copper tape
to obstruct potential infrared leakage paths.

\subsection{Dilution refrigeration and SQUID readout}
\label{sec:cryogenic_readout}

\begin{figure}[htbp]
\centering
\includegraphics[width=\linewidth]{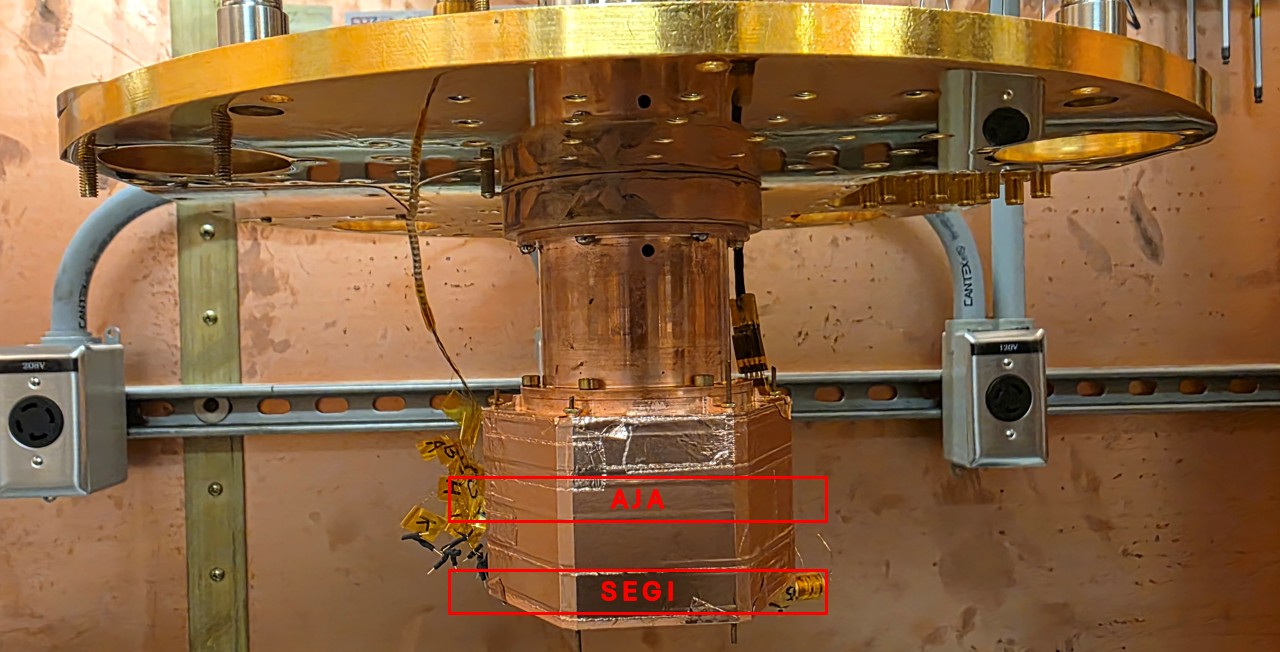}

\caption{
Cryogenic detector assembly mounted beneath the
mixing-chamber (MXC) plate of the Bluefors dilution
refrigerator. The AJA detector occupies the lower
position in the copper tower, and the SEGI detector
occupies the upper position.
}
\label{fig:tower}
\end{figure}

The copper tower was mounted beneath the mixing-chamber
(MXC) plate of a Bluefors dilution refrigerator, which
reached a measured base temperature of approximately
\SI{15}{\milli\kelvin}. The detector arrangement and
mounting are shown in Fig.~\ref{fig:tower}.

Channels B and C of each detector were connected to
superconducting quantum interference device (SQUID)
readout circuits, with channel C serving as the primary
measurement channel and channel B as a backup.
After cooldown, the SQUIDs were tuned for electrical
characterization and pulse acquisition.

\section{Electrical Characterization and Data Acquisition}
\label{sec:measurements}

\subsection{Superconducting switching and transition characterization}
\label{sec:electrical_characterization}

Electrical responses were monitored using a PicoScope
4824 at a display timebase of \SI{1}{\milli\second}
per division and a sampling rate of \(10\,\mathrm{MS/s}\),
corresponding to a sample interval of
\SI{100}{\nano\second}.

Following SQUID tuning, the channel bias was varied
to identify the onset of resistive behavior.
The corresponding current settings, reported here
as switching thresholds, were approximately
\SI{16.1}{\micro\ampere} for channels B and C of the
AJA detector and \SI{32}{\micro\ampere} for the
corresponding SEGI channels. Subsequent pulse
acquisition used bias settings at these measured
thresholds.

The superconducting transitions were investigated
by gradually warming the mixing-chamber (MXC) plate
using the Bluefors refrigerator's built-in heater.
A continuous triangular excitation with a generator
amplitude of \SI{10}{\milli\volt} peak-to-peak was
applied through the channel excitation circuit,
and the peak-to-peak SQUID-readout response was
monitored using the oscilloscope.

During warming, a pronounced decrease in the readout
amplitude was observed, consistent with redistribution
of the excitation current from the TES branch to the
parallel shunt resistor as the TES became resistive.
This decrease served as an operational indicator
of the superconducting transition.

Local temperatures were monitored during the sweeps
using ruthenium oxide resistive thermometers positioned
near each detector. Both thermometers had a reported
calibration accuracy of \SI{5}{\milli\kelvin} at
temperatures below \SI{150}{\milli\kelvin}.

%The readout-amplitude decreases occurred at approximately
The superconducting transition temperatures for the W TES were measured to be approximately
\SI{29}{\milli\kelvin} for the AJA detector and
\SI{37}{\milli\kelvin} for the SEGI detector. We believe that annealing of the tungsten TES during
the \SI{150}{\degreeCelsius} preparation step before
aluminum deposition in the AJA system reduced
\(T_{\mathrm{c}}\) by a few millikelvin.
%These values can be reported as transition temperatures for each sample.

\subsection{Pulse acquisition}
\label{sec:pulse_acquisition}

Following electrical characterization, the detectors
were biased at the measured switching thresholds
reported in Section~\ref{sec:electrical_characterization}.
Channel C of each detector served as the primary
measurement channel, with channel B available as
a backup.

A \(^{22}\mathrm{Na}\) radioactive source was placed
outside the dilution refrigerator, on the vacuum can,
to increase the event rate for the detector-response
comparison. The source produces \SI{511}{\kilo\electronvolt}
annihilation photons and an approximately
\SI{1274.5}{\kilo\electronvolt} gamma ray from
de-excitation of \(^{22}\mathrm{Na}\).
Given the external source placement and intervening
refrigerator materials, identifiable full-energy peaks
were not expected in the measured spectra.
The source was therefore used to enhance event
statistics rather than to provide an energy calibration.

Waveforms were acquired using a PicoScope 4824
with a \SI{10}{\milli\second} record window and
a 20\% pretrigger fraction, corresponding to
\SI{2}{\milli\second} of pretrigger data.
The trigger threshold was \SI{30}{\milli\volt}
at the oscilloscope input, and the nominal ADC
resolution was 12 bits. The recorded waveforms
were subsequently analyzed to construct
uncalibrated pulse-estimator spectra.

\section{Analysis Methods}
\label{sec:analysis}

\subsection{Pulse templates and optimal-filter analysis}
\label{sec:optimal_filter}

\begin{figure}[htbp]
\centering
\includegraphics[width=\linewidth]{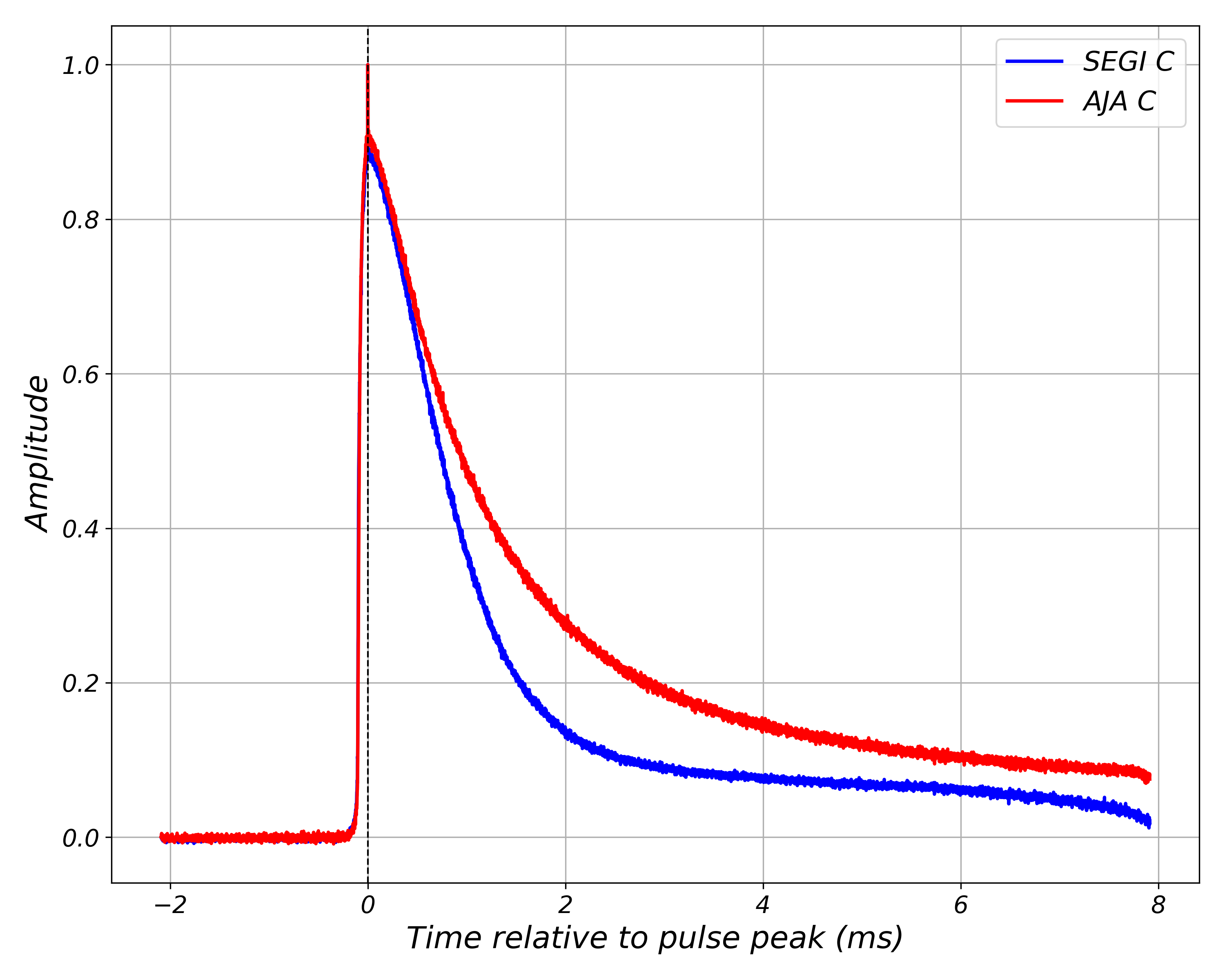}
\caption{
Pulse templates constructed from the recorded
channel C waveforms of the AJA and SEGI detectors
and used in the optimal-filter analysis.
}
\label{fig:template}
\end{figure}

Pulse templates were constructed from the recorded
channel C waveforms of each detector and used in
an optimal-filter analysis to estimate event amplitudes.
The templates shown in Fig.~\ref{fig:template}
exhibit a longer decay tail for AJA than for SEGI,
indicating different temporal responses. These
differences may reflect quasiparticle collection,
energy-transfer dynamics, or TES electrothermal
response.

\begin{figure}[ht!]
\centering
\includegraphics[width=\linewidth]{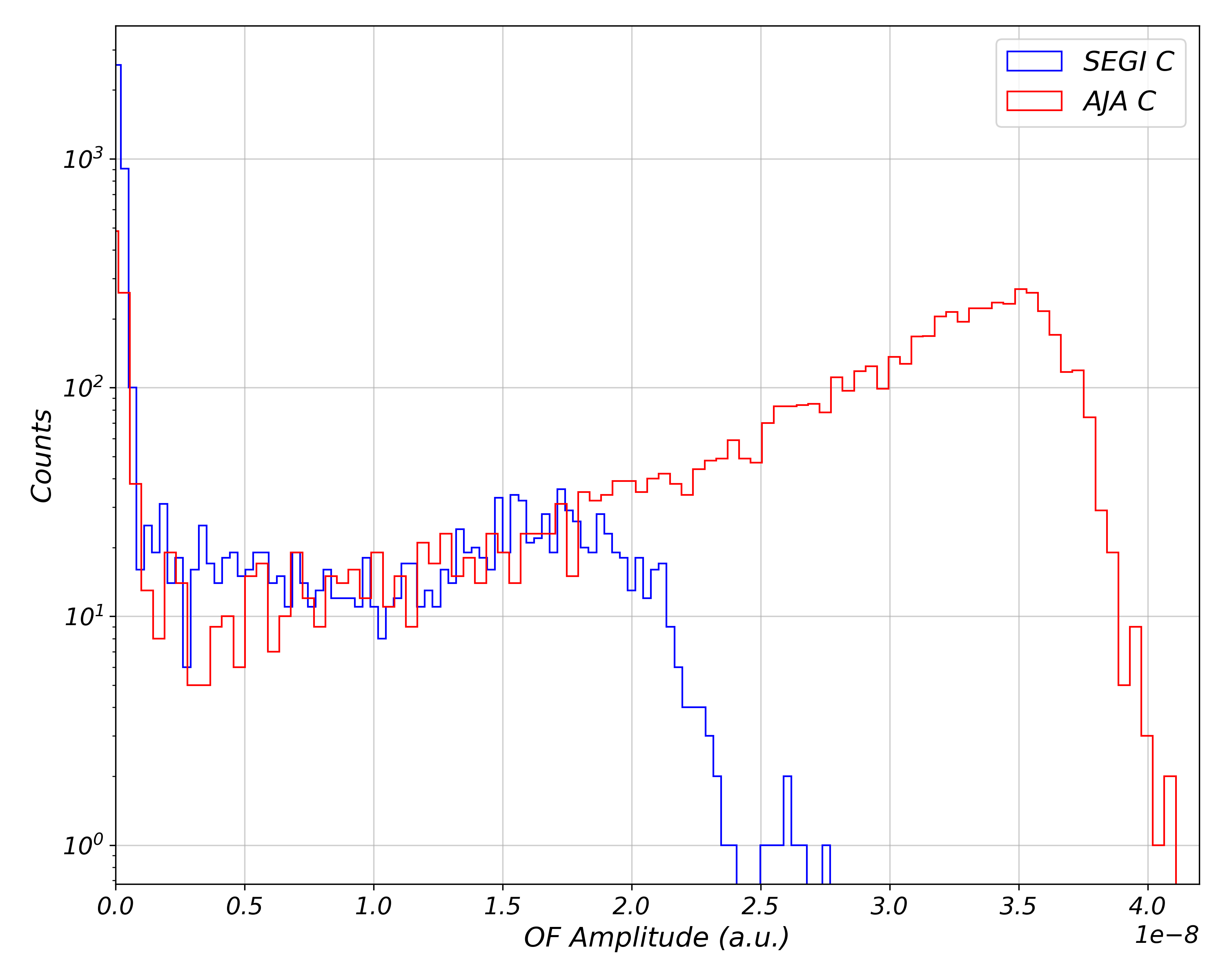}

\caption{
Uncalibrated optimal-filter pulse-amplitude spectra
for channel C of the AJA and SEGI detectors. The AJA spectrum is shifted
toward larger reconstructed amplitudes,
indicating a larger measured signal response.
This observation is consistent with improved
quasiparticle energy transfer to the tungsten
TES. It shows the combined effect of quasiparticle transport and the transmission of the quasiparticles across the Al/W interface.
}
\label{fig:pulse_spectra}
\end{figure}

The amplitudes reconstructed using the optimal filter
were histogrammed to produce the uncalibrated
pulse-estimator spectra for the AJA and SEGI detectors
shown in Fig.~\ref{fig:pulse_spectra}.

\begin{figure}[ht!]
\centering
\includegraphics[width=\linewidth]{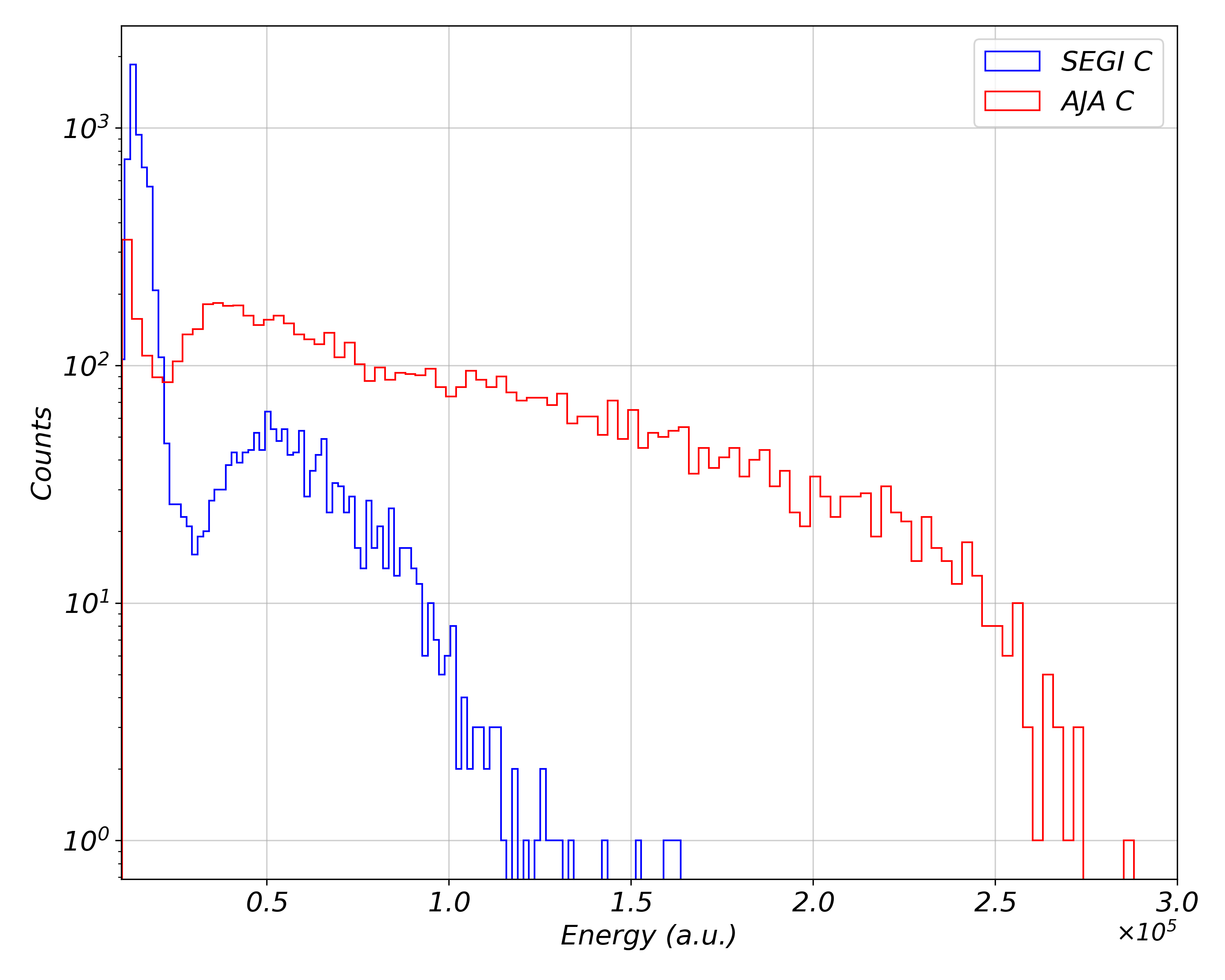}
\caption{
    Pulse-integral spectra constructed using a
    \SI{6}{\milli\second} integration window for
    channel C waveforms of the AJA and SEGI detectors,
    illustrating the contribution of the longer
    AJA pulse tails.
}
\label{fig:raw spectra}
\end{figure}

To examine the contribution of the longer tail,
pulse-integral spectra were also constructed using
a \SI{6}{\milli\second} integration window, as shown
in Fig.~\ref{fig:raw spectra}.
Increasing the integration window extended the
AJA spectrum toward larger pulse-integral values,
whereas the SEGI spectrum remained approximately
unchanged. This behavior is consistent with a
greater contribution from the late-time AJA response.

\subsection{Interpretation of the spectral comparison}
\label{sec:response_comparison}

The spectral comparison was motivated by the
expectation that more efficient transfer of
quasiparticle energy from the aluminum collectors
to the tungsten TES elements would increase
the measured response to a comparable energy
deposition. Under otherwise comparable detector
and readout conditions, this increase could
shift corresponding features in the
pulse-estimator spectrum toward larger
reconstructed amplitudes. The QET structure with higher grained aluminum (and higher RRR) leads to longer inelastic diffusion lengths and extended recombination lifetimes within the superconducting fins. Consequently, it takes longer for the quasiparticles to propagate to and get trapped by the tungsten Transition Edge Sensor, spreading the thermalization and energy collection out over a longer duration. Such a shift is
observed in Figs.~\ref{fig:pulse_spectra} and \ref{fig:raw spectra}.

\section{Results}
\label{sec:results}

The AJA aluminum witness samples exhibited a mean
room-temperature-to-\SI{10}{\kelvin} resistance ratio
of approximately 18, compared with 9 for SEGI,
based on three samples per deposition process
(Section~\ref{sec:rrr}).
AFM measurements of the separate characterization
films also revealed greater grain size
for AJA: $R_q=\SI{132}{\nano\metre}$ and
$R_a=\SI{112}{\nano\metre}$, compared with
$R_q=\SI{9.54}{\nano\metre}$ and
$R_a=\SI{5.60}{\nano\metre}$ for SEGI
(Tab.~\ref{tab:afm_results} and
Fig.~\ref{fig:afm_characterization}).

Electrical characterization yielded switching
thresholds of approximately
\SI{16.1}{\micro\ampere} for AJA channels B and C
and \SI{32}{\micro\ampere} for the corresponding
SEGI channels. Mean transition temperatures
inferred from repeated readout-response measurements
were approximately \SI{29}{\milli\kelvin} for AJA
and \SI{37}{\milli\kelvin} for SEGI, as measured
by the nearby thermometer
(Section~\ref{sec:electrical_characterization}).

The principal detector-response difference was
the displacement of the AJA optimal-filter spectrum
toward larger amplitudes
(Fig.~\ref{fig:pulse_spectra}).
The median
amplitudes were $3.454\times10^{-8}$~a.u. for
AJA channel C and $1.639\times10^{-8}$~a.u. for
SEGI channel C. Their ratio was 2.108, with a
95\% paired-bootstrap confidence interval of
$[2.060,\,2.156]$.
This comparison characterizes the selected,
uncalibrated response distributions rather than
an absolute improvement in collected energy or
collection efficiency. Measurements of these
quantities are ongoing and will be reported
in future work.

\section{Discussion}
\label{sec:discussion}

The AJA deposition process produced aluminum witness
samples with approximately twice the RRR of the SEGI
samples and a detector with a larger uncalibrated
optimal-filter response. This association is consistent
with improved quasiparticle-energy transport or transfer
at the Al/W interface \cite{Yen2014}, but does not
establish either mechanism. RRR does not directly
measure quasiparticle lifetime or collection efficiency.

Future work will investigate quasiparticle transport
using TES-based measurements of energy sharing in
superconducting aluminum films \cite{Yen2014}.
Within a diffusion model with a first-order loss term,
fitting these measurements will provide an effective
diffusion length, \(L_{\mathrm{diff}}\). Combined with
an independently measured diffusion constant, \(D\),
this yields an effective quasiparticle loss time,
\begin{equation}
\tau_{\mathrm{qp,eff}}
=
\frac{L_{\mathrm{diff}}^{2}}{D}.
\label{eq:qp_lifetime}
\end{equation}
This effective time need not equal the intrinsic
recombination lifetime. Complementary measurements
will use coplanar waveguide resonators fabricated
from high-RRR AJA-deposited aluminum to determine
quasiparticle lifetimes from the resonator dynamics,
with temperature and microwave readout power
controlled.

\section{Conclusion}
\label{sec:conclusion}

Two silicon phonon detectors with
a common nominal four-channel layout
were fabricated to compare aluminum
deposition using AJA and SEGI
sputtering systems. Both detectors
used \SI{1}{\milli\metre}-thick
silicon substrates and nominally
\SI{40}{\nano\metre}-thick tungsten
TES films deposited using SEGI.
The patterned tungsten structures
were subsequently coupled to
nominally \SI{600}{\nano\metre}-thick
aluminum collectors.

The aluminum witness samples exhibited
room-temperature-to-\SI{10}{\kelvin}
resistance ratios of approximately
18 for AJA and 9 for SEGI.
AFM measurements on separate
\SI{300}{\nano\metre} characterization
films showed greater surface roughness
for the AJA sample, which directly shows the grain size. % demonstrating that the higher reported RRR was not accompanied by a smoother measured surface, otherwise, it showed bigger grains.

The TES biasing procedure yielded operating currents
of approximately \SI{16.1}{\micro\ampere} for the AJA
detector and \SI{32}{\micro\ampere} for the SEGI
detector at the steepest part of the tungsten TES
transition. Operational transition
temperature estimates were approximately
\SI{29}{\milli\kelvin} and
\SI{37}{\milli\kelvin}, respectively. The lower bias point accurately reflects the lower transition temperature of the AJA W TES, since the bath temperature was approximately 15 mK. The slower pulse fall time is expected to be dominated by the continuous flow of quasiparticles from the aluminum fins to the TES over a longer period due to the higher quasiparticle lifetime, rather than the small Tc difference between the two devices.

Under \(^{22}\mathrm{Na}\) irradiation,
the channel C optimal-filter spectrum
of the AJA detector was shifted
toward larger reconstructed amplitudes.
This establishes a larger measured
uncalibrated response for the AJA
device under the conditions of
this comparison. The observation is consistent with
improved delivery of quasiparticle
energy to the tungsten TES.

The present comparison identifies
the AJA-based fabrication procedure
as a promising candidate for further
optimization of aluminum collectors
and Al/W interfaces in QET-based
silicon phonon detectors.

Motivated by the observed response improvement and
the inferred increase in effective quasiparticle
lifetime, we aim to investigate larger QET geometries. The achievable QET area scales as the square of
the improvement factor in quasiparticle lifetime. Thus, with the improvement presented here we can effectively make our QETs four times larger in area for a vastly superior phonon signal sensitivity.

We aim to measure the quasiparticle lifetime directly
using techniques already employed within our broader
collaborations~\cite{Yen2014}, as well as techniques
based on microwave resonators~\cite{PhysRevLett.100.257002}.

\section{Acknowledgements}
This work was supported by the U.S. DOE (DE-SC0018981, DE-SC0010813) and the Mitchell Institute at Texas A\&M University.
%The authors gratefully acknowledge the financial support of the Mitchell Institute for Fundamental Physics and Astronomy, as well as DOE funds ..... 
A portion of the nanofabrication work was conducted in the Texas A\&M University AggieFab Nanofabrication Facility (RRID:SCR023639), which is supported by the Texas A\&M Engineering Experiment Station and Texas A\&M University.

\begingroup
\raggedright
\bibliographystyle{elsarticle-num}
\bibliography{ref}

@article{Irwin1995,
  author  = {Irwin, K. D. and Nam, S. W. and Cabrera, B.
             and Chugg, B. and Young, B. A.},
  title   = {A quasiparticle-trap-assisted transition-edge sensor
             for phonon-mediated particle detection},
  journal = {Review of Scientific Instruments},
  volume  = {66},
  number  = {11},
  pages   = {5322--5326},
  year    = {1995},
  doi     = {10.1063/1.1146105}
}

@article{Yen2014,
  author  = {Yen, J. J. and Shank, B. and Young, B. A.
             and Cabrera, B. and Brink, P. L. and Cherry, M.
             and Kreikebaum, J. M. and Moffatt, R.
             and Redl, P. and Tomada, A. and Tortorici, E. C.},
  title   = {Measurement of quasiparticle transport in aluminum
             films using tungsten transition-edge sensors},
  journal = {Applied Physics Letters},
  volume  = {105},
  number  = {16},
  pages   = {163504},
  year    = {2014},
  doi     = {10.1063/1.4899130}
}

@article{Jastram:2014pga,
    author = "Jastram, A. and Harris, H. R. and Mahapatra, R. and Phillips, J. and Platt, M. and Prasad, K. and Sander, Joel and Upadhyayula, S.",
    title = "{Cryogenic Dark Matter Search Detector Fabrication Process and Recent Improvements}",
    eprint = "1408.0295",
    archivePrefix = "arXiv",
    primaryClass = "physics.ins-det",
    doi = "10.1016/j.nima.2014.10.043",
    journal = "Nucl. Instrum. Meth. A",
    volume = "772",
    pages = "14--25",
    year = "2015",
    note = "[Erratum: Nucl.Instrum.Meth.A 820, 172--172 (2016)]"
}

@article{CDMS-II:2009ktb,
    author = "Ahmed, Z. and others",
    collaboration = "CDMS-II",
    title = "{Dark Matter Search Results from the CDMS II Experiment}",
    eprint = "0912.3592",
    archivePrefix = "arXiv",
    primaryClass = "astro-ph.CO",
    reportNumber = "FERMILAB-PUB-09-637-AE",
    doi = "10.1126/science.1186112",
    journal = "Science",
    volume = "327",
    pages = "1619--1621",
    year = "2010"
}

@article{article,
author = {Brink, P. and Anderson, Adam and Balakishiyeva, D. and Bauer, D. and Beaty, J. and Brandt, Daniel and Cabrera, Blas and Chagani, H. and Cherry, Minn and Cooley, J. and Silva, E. and Cushman, Priscilla and Daal, Miguel and Doughty, Todd and Figueroa-Feliciano, Enectali and Fritts, M. and Godfrey, Gary and Golwala, S. and Hall, Jeter and Zhang, Jujia},
year = {2014},
month = {02},
pages = {},
title = {Detector Fabrication Yield for SuperCDMS Soudan},
volume = {176},
journal = {Journal of Low Temperature Physics},
doi = {10.1007/s10909-014-1100-1}
}

@phdthesis{BihariPrasad:2013mfw,
    author = "Bihari Prasad, Kunj",
    title = "{Search for lightly ionizing particles using CDMS-II data and fabrication of CDMS detectors with improved homogeneity in properties}",
    reportNumber = "FERMILAB-THESIS-2013-32",
    doi = "10.2172/1155185",
    school = "Texas A-M",
    year = "2013"
}

@ARTICLE{546406,
  author={Williams, K.R. and Muller, R.S.},
  journal={Journal of Microelectromechanical Systems}, 
  title={Etch rates for micromachining processing}, 
  year={1996},
  volume={5},
  number={4},
  pages={256-269},
  doi={10.1109/84.546406}}

@article{PhysRevLett.20.1502,
  title = {Diffusion of Quasiparticles in Superconducting Aluminum Films},
  author = {Hsieh, S. Y. and Levine, James L.},
  journal = {Phys. Rev. Lett.},
  volume = {20},
  issue = {26},
  pages = {1502--1504},
  numpages = {0},
  year = {1968},
  month = {Jun},
  publisher = {American Physical Society},
  doi = {10.1103/PhysRevLett.20.1502},
  url = {https://link.aps.org/doi/10.1103/PhysRevLett.20.1502}
}

@article{LOIDL2001440,
title = {Quasiparticle diffusion over several mm in cryogenic detectors},
journal = {Nuclear Instruments and Methods in Physics Research Section A: Accelerators, Spectrometers, Detectors and Associated Equipment},
volume = {465},
number = {2},
pages = {440-446},
year = {2001},
issn = {0168-9002},
doi = {https://doi.org/10.1016/S0168-9002(01)00621-0},
url = {https://www.sciencedirect.com/science/article/pii/S0168900201006210},
author = {M. Loidl and S. Cooper and O. Meier and F. Pröbst and G. Sáfrán and W. Seidel and M. Sisti and L. Stodolsky and S. Uchaikin}
}

@article{TESSERACT:2025odn,
    author = "Chang, C. L. and others",
    collaboration = "TESSERACT",
    title = "{Spontaneous generation of athermal phonon bursts within bulk silicon causing excess noise, low energy background events, and quasiparticle poisoning in superconducting sensors}",
    eprint = "2505.16092",
    archivePrefix = "arXiv",
    primaryClass = "physics.ins-det",
    doi = "10.1063/5.0281876",
    journal = "Appl. Phys. Lett.",
    volume = "127",
    number = "26",
    pages = "263502",
    year = "2025"
}

@article{TESSERACT:2025tfw,
    author = "Bui, T. K. and others",
    collaboration = "TESSERACT",
    title = "{First Limits on Light Dark Matter Interactions in a Low Threshold Two-Channel Athermal Phonon Detector from the TESSERACT Collaboration}",
    eprint = "2503.03683",
    archivePrefix = "arXiv",
    primaryClass = "hep-ex",
    doi = "10.1103/hsrl-crvf",
    journal = "Phys. Rev. Lett.",
    volume = "135",
    number = "16",
    pages = "161002",
    year = "2025"
}

@article{PhysRevLett.100.257002,
  title = {Quasiparticle Relaxation in Optically Excited High-$Q$ Superconducting Resonators},
  author = {Barends, R. and Baselmans, J. J. A. and Yates, S. J. C. and Gao, J. R. and Hovenier, J. N. and Klapwijk, T. M.},
  journal = {Phys. Rev. Lett.},
  volume = {100},
  issue = {25},
  pages = {257002},
  numpages = {4},
  year = {2008},
  month = {Jun},
  publisher = {American Physical Society},
  doi = {10.1103/PhysRevLett.100.257002},
  url = {https://link.aps.org/doi/10.1103/PhysRevLett.100.257002}
}

@article{MINER:2025wrk,
    author = "Mondal, D. and others",
    collaboration = "MINER",
    title = "{CE{\ensuremath{\nu}}NS search with cryogenic sapphire detectors at MINER: Results from the TRIGA reactor data and future sensitivity at HFIR}",
    eprint = "2510.15999",
    archivePrefix = "arXiv",
    primaryClass = "nucl-ex",
    doi = "10.1103/115m-y8mg",
    journal = "Phys. Rev. D",
    volume = "114",
    number = "3",
    pages = "032009",
    year = "2026"
}

@article{MINER:2026sat,
    author = "Mondal, D. and others",
    collaboration = "MINER",
    title = "{Exploring low energy excess in MINER with sapphire detectors using a convolutional variational autoencoder}",
    eprint = "2605.31190",
    archivePrefix = "arXiv",
    primaryClass = "physics.ins-det",
    doi = "10.1103/21cx-xmgh",
    journal = "Phys. Rev. D",
    volume = "114",
    number = "5",
    pages = "052015",
    year = "2026"
}
\endgroup

\end{document}